\documentclass[superscriptaddress,prb,twocolumn,floatfix]{revtex4-2}

\usepackage{graphicx}
\usepackage[caption=false]{subfig}
\usepackage{siunitx}
\DeclareSIUnit\mub{\ensuremath{\mu_B}}
\DeclareSIUnit\formulaunit{\text{f.u.}}
\DeclareSIUnit\angstrom{\text{\AA}}
\DeclareSIUnit\elementarycharge{\ensuremath{e}}
\DeclareSIUnit\permille{\text{\textperthousand}}
\DeclareSIUnit\atomicunit{\text{a.u.}}
\DeclareSIUnit\rydberg{\text{Ry}}
\usepackage{booktabs}
\usepackage{amssymb}

\newcommand{\tcmp}{\ensuremath{T_{\text{comp}}}}
\newcommand{\tdep}{\ensuremath{T_{\text{dep}}}}
\newcommand{\mra}{\ensuremath{\text{Ru}_{2-x}\text{Mn}_{1+x}\text{Al}}}
\newcommand{\mrx}[1]{\ensuremath{\text{Ru}_{\num[evaluate-expression = true, round-mode = places,
round-precision = 1]{2 - #1}}\text{Mn}_{\num[evaluate-expression = true, round-mode = places,
round-precision = 1]{1 + #1}}\text{Al}}}

\begin{document}
\title{\mra{} thin films}

\author{K.E.\,Siewierska}
\email{siewierk@tcd.ie}
\affiliation{School of Physics and CRANN, Trinity College Dublin, Ireland}
\author{H.\,Kurt}
\affiliation{Istanbul Medeniyet University, Department of Engineering Physics, Goztepe Istanbul 34700, Turkey
}%
\author{B.\,Shortall}
\author{A.\,Jha}
\author{N.\,Teichert}
\author{G.\,Atcheson} 
\author{M.\,Venkatesan} 
\author{J.M.D.\,Coey}
\author{Z.\,Gercsi}
\author{K.\,Rode}
\affiliation{School of Physics and CRANN, Trinity College Dublin, Ireland}%

\begin{abstract}
  The cubic Heusler alloy \mra{} is grown in thin film form on MgO and
  MgAl$_2$O$_4$
  substrates. It is a highly spin-polarised ferrimagnetic metal, with
  weak magnetocrystalline anisotropy. Although structurally and
  chemically similar to $\text{Mn}_2\text{Ru}_x\text{Ga}$, it does not
  exhibit ferrimagnetic compensation, or large magneto galvanic effects.
  The differences are attributed to a combination of atomic order
  and the hybridisation with the group 13 element Al or Ga. The spin
  polarisation is around \qtyrange{50}{60}{\percent}. There is a gap in the
  density of states just above the Fermi level in fully ordered compounds.
\end{abstract}

\maketitle

\section{\label{sec:intro}Introduction}
Antiferromagnetic spin electronics is attracting considerable interest due to
the relative abundance of antiferromagnetically coupled materials, the
fast magnetisation dynamics and the insensitivity to
external magnetic fields along with the absence of demagnetising forces. A
major difficulty is control and read-out of the antiferromagnetic state. A
compensated ferrimagnet where two antiferromagnetically coupled,
but inequivalent, sublattices produce zero net moment at the compensation
temperature \tcmp{} potentially combines the advantages of
antiferromagnets,\cite{Wollmann2017,GRAF20111} with those of a
ferromagnetic metal. The magnetisation can be controlled by
an external field away from \tcmp{}, coupled with a high transport spin
polarisation that facilitates reading the magnetic state using magneto-optical
Kerr effect\cite{Teichert2020,Siewierska_MOKE,banerjee2019single}, anomalous
Hall effect (AHE),\cite{Thiyagarajah.2015} and giant- or
tunnel-\-magneto\-resistance\cite{Borisov.2016} (GMR or TMR).

In \citeyear{vanLeuken.1995}, \citet{vanLeuken.1995} proposed that some
half-Heusler alloys (C1$_{\text{b}}$ structure) would exhibit magnetic
compensation,
yet due to the inequivalent magnetic sublattices exhibit \qty{100}{\percent}
spin polarisation and thus be half metallic. Subsequent work followed the empirical
Slater-Pauling rule,\cite{Galanakis} $M = N_v - 18$ for half-Heuslers and $M =
N_v - 24$ for full Heuslers. $N_v$ is the number
of valence electrons, $M$ the magnetisation in $\unit{\mub\text{ per formula
unit}}$ (\unit{\mub\per\formulaunit}). The D0$_3$ phase of
Mn$_3$Ga and Mn$_3$Al were suggested as possible compensated half-metallic ferrimagnets,\cite{Wurmehl}
but neither crystallize in the D0$_3$
structure in the bulk. There is a report on the growth of D0$_3$-structure
Mn$_3$Al thin films on GaAs substrates with a net moment of
\qty{0.017}{\mub\per\formulaunit} and a Curie temperature of
\qty{605}{\kelvin}\cite{Jamer.2017}. Disordered antiferromagnetic Mn$_3$Al
films with a cubic structure and a Ne\'{e}l temperature of \qty{400}{\kelvin}
were also grown on MgO substrates\cite{Bang2019}. The equilibrium phase for
Mn$_2$Ga and Mn$_3$Ga is hexagonal D0$_{19}$ which can be changed to
tetragonal D0$_{22}$ by annealing at
\qty{400}{\degreeCelsius}.\cite{Kurtpssb2011} On the other hand Mn$_2$Al and
Mn$_3$Al crystallize in the cubic $\beta$-Mn structure and do not order
magnetically but are spin glasses.\cite{RuiZhangJMMM2020} The materials can
also be grown in the D0$_{19}$ hexagonal structure when they are produced as
epitaxial sputtered films on suitable seed layers\cite{Liu1999}.

The first experimental example of a compensated half metallic ferrimagnet,
was MRG, the near-cubic Mn-Ru-Ga alloy with formula Mn$_2$Ru$_{x}$Ga, discovered by \citet{Kurt.2014} in
\citeyear{Kurt.2014}. MRG crystallises in space group $F\bar{4}3m$ (XA).  The
Ru concentration $x$
allows tuning of \tcmp{} and also helps to stabilise the near-cubic structure.
The original composition with $x = 0.5$ was initially thought to have $N_v =
21$ valence electrons and a half-filled Ru sublattice, but a recent
study\cite{Siewierska2021_PRB} established that the films contained few vacant
sites and $N_v \approx 24$.
Compensated ferrimagnetism has also been demonstrated in bulk
Mn$_{1.5}$FeV$_{0.5}$Al, a quaternary Heusler alloys with 24 valence
electrons\cite{Stinshoff2017,Stinshoff2017_2}. Half-metallic compensated
ferrimagnetic materials, such as MRG, are of interest for spin orbit torque
switching\cite{Finley.2019,Finley2020} as well as magnetic oscillations in the
terahertz region for high-speed,\cite{banerjee2021,DaviesPRR2020} on-chip communications.\cite{Betto.2016,Chen2018}.

The magnetic and transport properties of MRG including their dependence on the
Ru content $x$ are understood in a rigid band model. One sublattice
(formed by states originating from Mn in Wyckoff position $4c$) dominates
the band structure around the Fermi level and is coupled antiferromagnetically
to another (formed by Mn in position $4a$) whose states are sufficiently deep
to not contribute significantly to the transport. In this model, the role of Ru
is to contribute extra electrons to the $4c$ sublattice and hence increase its
moment and thus \tcmp{}. The role of the group 13 element, Ga, is ignored.
Albeit simplistic, this model explains most of the properties of
MRG\@. Its experimental validation was provided by \citet{Siewierska2021_PRB}
who found an excellent linear relation between the number of valence electrons and
magnetisation -- one added valence electron increases the magnetisation by
one \unit{\mub\text{ per formula unit}}.

Here we change from Ga to Al to produce thin films of MRA, the Mn-Ru-Al Heusler
alloy whose formula is best written \mra{}, and
report optimised growth conditions, structural and magnetic properties along
with magnetotransport and spin polarisation measurements for films crystallizing
in space group $Fm\bar{3}m$ (L2$_1$ structure) with Al and Mn in the $4a$-$4b$
plane and Ru occupying the $8c$ central cube. When $x > 0$, excess Mn occupies
a fraction of the $8c$ positions, now split into $4c$ and $4d$. Note that $4a,c$
and $4b,d$ are symmetrically equivalent. The comparison of the
two compounds illustrates the importance of crystalline order in this group of
Heusler alloys and highlights the importance of Ga.

\section{\label{sec:methods}Methodology}
Epitaxial thin films of MRA were grown by DC magnetron sputtering, using our
Shamrock sputtering system, on \qtyproduct{10x10}{\milli\metre} (100) SrTiO$_3$
(STO), MgAl$_2$O$_4$ (MAO) and MgO substrates.  Samples were co-sputtered in
Argon from a Mn$_2$Al target (from Kojundo Chemical Laboratory Co. Ltd, Japan)
and a Ru target in a confocal sputtering geometry onto the substrates maintained
at an optimized deposition temperature (\tdep{}) we found to be
\qty{425}{\degreeCelsius}. Prior to deposition, the back of the substrates was coated
with Ti or Ta to ensure uniform absorption of heat and keep \tdep{}
constant during deposition. The films had an average thickness of
\qty{60}{\nano\metre} and were capped \emph{in-vacuo} with a
\qty{2}{\nano\metre} layer of AlO$_x$ deposited at room temperature in order to
prevent oxidation. We note that this arrangement of targets does not yield
perfect \mra{} stoichiometry as a decrease in Mn also leads to
decreased Al content. The sum of the amounts of Ru and Mn in the formula unit
varies from \numrange{3.0}{3.3} in our samples, and we therefore write the
formula as \mra{} where it is implicit that Al is sub-stoichiometric by up to
\qty{30}{\percent}.

A Bruker D8 Discover X-ray diffractometer with a copper K$\alpha$
X-rays and a double-bounce Ge [220] monochromator ($\lambda =
\qty{154.06}{\pico\metre}$) was used to determine the diffraction patterns and
reciprocal space maps (RSM) of the thin films. The substrate (113) reflection
is in the same plane as the MRA (204) reflection and it was used to calculate the
lattice parameters. Low angle X-ray reflectivity and symmetric X-ray
diffraction patterns were measured using a Panalytical X'Pert Pro ($\lambda =
\qty{154.19}{\pico\metre}$) diffractometer. Film thickness and density were
found by fitting the interference pattern using X'Pert Reflectivity software.
The densities were used to deduce the stoichiometric ratios assuming full
site occupation.\cite{Siewierska2021_PRB}

Low-field anomalous Hall effect (AHE) measurements were performed in a \qty{1}{\tesla} GMW
electromagnet system in ambient conditions. Silver wires were cold-welded to
the thin films with indium and the current used was \qty{5}{\milli\ampere}.
Higher field measurements were performed in a superconducting magnet which had
a maximum field of \qty{4}{\tesla} with a cryostat for low-temperature
magneto-transport. The films were contacted with silver paint. The 4-point Van
der Pauw geometry was used to determine both the Hall and the longitudinal
resistivity of the films. 

Point contact Andreev reflection (PCAR) measurements were also performed in a
Quantum Design PPMS using a mechanically-sharpened Nb tip. The landing of tip
onto the sample surface is carefully controlled by an automated vertical
attocube\textsuperscript{TM} piezo stepper. Two horizontal steppers are used
to move the sample laterally to probe a pristine area. The differential
conductance spectra were fitted using a modified Blonder-Tinkham-Klapwijk
model, as detailed elsewhere.\cite{PCAR_1,PCAR_2}

Measurements of the magnetisation with a magnetic field applied perpendicular
or parallel to the surface of the films were carried out using a
\qty{5}{\tesla} Quantum Design SQUID magnetometer These data were corrected for the
diamagnetism of the substrate by linear subtraction.

Ab-initio calculations based on density functional theory (DFT) were carried
out using norm-conserving pseudopotentials and pseudo-atomic localized basis
functions implemented in the OpenMX software package.\cite{openmx} The generalized
gradient approximation (GGA-PBE) approach was used for all calculations.
The structure was fully relaxed to minimize interatomic forces. We used a
16-atom supercell cell with \numproduct{17 x 17 x 17} $k$-points to evaluate the total energies.
Pre-generated pseudopotentials and pseudo-atomic orbitals with a cut-off radius
of 6, 7 and 7 atomic units (\unit{\atomicunit}) were used for Mn, Ru and Al elements,
respectively.  An energy cut-off of \qty{300}{\rydberg} was used for numerical integrations.
The convergence criterion for the energy minimization procedure was set to
\qty{e-8}{Hartree}. The spin orbit interaction (SOI) was turned off for the calculations.

\section{\label{sec:results}Results \& Discussion}
\subsection{\label{ssec:structure}Structure}
We first determine a suitable substrate for the growth of \mra{} thin films,
and optimise the deposition temperature \tdep{}. MRG has a lattice
parameter of $a_0 \approx \qty{600}{\pico\metre}$ and its epitaxial relation with
the MgO substrate ensures that the in-plane $[100]$ direction of MgO is
parallel to the $[110]$ direction of MRG, $a_{\text{MRG}} \approx \sqrt{2}
a_{\text{MgO}} = \qty{595.6}{\pico\metre}$. We therefore explore the growth of
MRA on
MgO, SrTiO$_3$ (STO, $\sqrt{2} a = \qty{552.3}{\pico\metre}$) and MgAl$_2$O$_4$
(MAO, $\sqrt{2} a = \qty{571.8}{\pico\metre}$).

\begin{figure}
  \centering
  \subfloat{\label{fig:xrd-1}\includegraphics{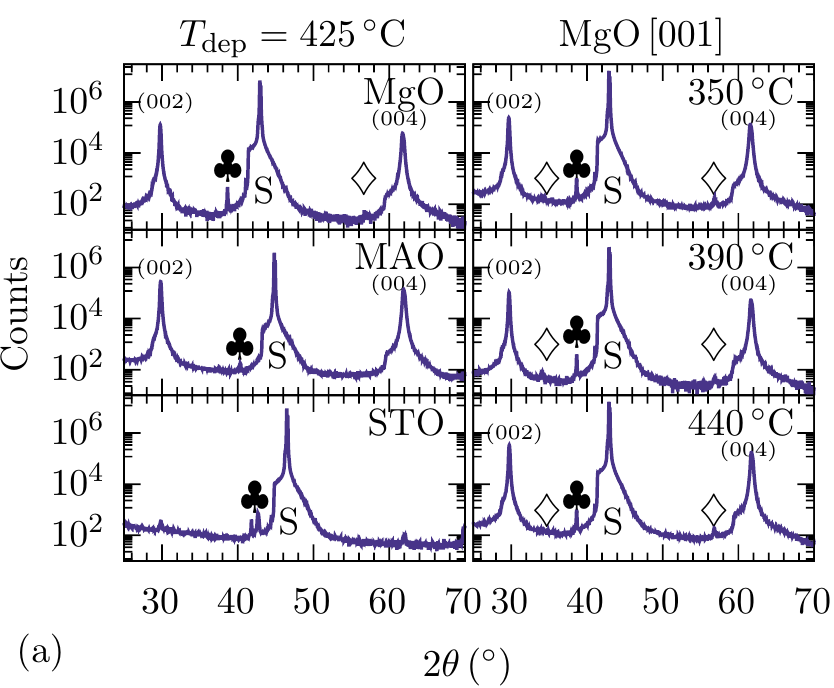}}\\
  \subfloat{\label{fig:xrd-2}\includegraphics{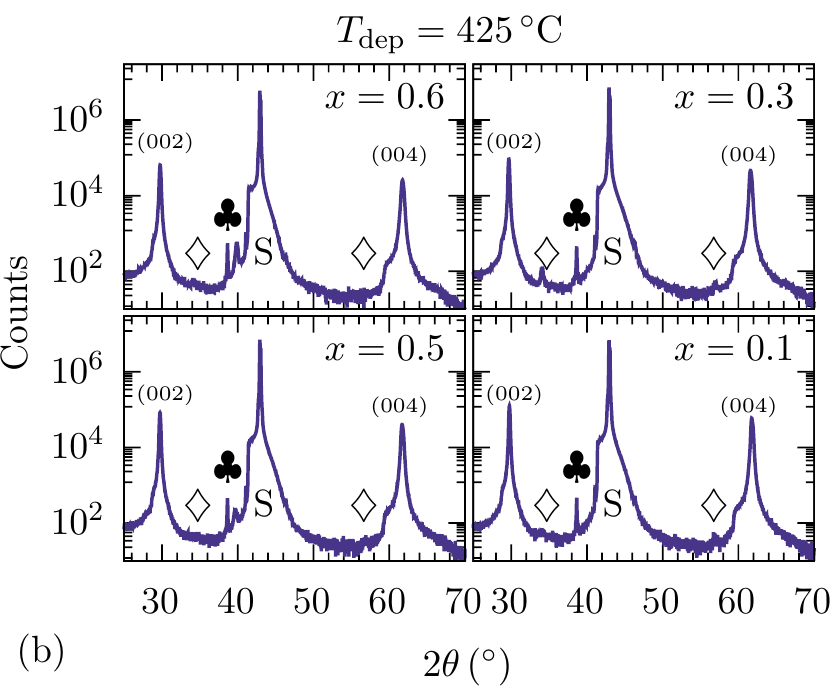}}
  \caption[]{\label{fig:xrd}X-ray diffraction patterns. \subref{fig:xrd-1}
    \mrx{0.10} deposited on MgO, MAO and STO substrates at $\tdep{} =
    \qty{425}{\degreeCelsius}$ (left column) and on MgO with varying \tdep{}
    (right column). \subref{fig:xrd-2} \mra{} deposited at $\tdep{} =
    \qty{425}{\degreeCelsius}$ for $x = \numlist{0.1;0.3;0.5;0.6}$. The
    substrate (002) reflection is marked by `S', while the corresponding Cu
    $K_{\beta}$ reflection ($\lambda = \qty{139.23}{\pico\metre}$) is marked by
    $\clubsuit$. We also indicate a minor secondary phase
    ($\diamondsuit$). 
  }%
\end{figure}
In \figurename~\ref{fig:xrd-1} (left column) we show X-ray diffraction
patterns of \mrx{0.10} on MgO, MAO and STO\@. MRA does not
crystallize on STO, while on both MAO and MgO the films are fully textured with the
growth axis parallel to the $[001]$ crystal direction. There are only minor
differences between MgO and MAO, and we select MgO as the
preferred substrate. In \figurename~\ref{fig:xrd-1} (right column) the
deposition temperature \tdep{} is varied from \qtyrange{350}{440}{\degreeCelsius}.
Although a minor secondary phase is present in all samples,  it is nearly
suppressed at $\tdep{} = \qty{425}{\degreeCelsius}$. Finally, in
\figurename~\ref{fig:xrd-2} we show the patterns for \mra{} films with $x =
\numlist{0.1;0.3;0.5;0.6}$. The films are fully textured in the entire composition range, 
with only a minor secondary phase 
whose associated peaks are about \numrange{3}{4} orders of
magnitude less intense than the MRA peaks. A likely origin is
a very small amount ($< \qty{1}{\percent}$) of either a Ru-Mn\cite{YIN201570}
or Ru-Al binary alloy in the film, or a Ru oxide at the interface between the
film and the capping layer.

\begin{figure}
  \centering
  \includegraphics{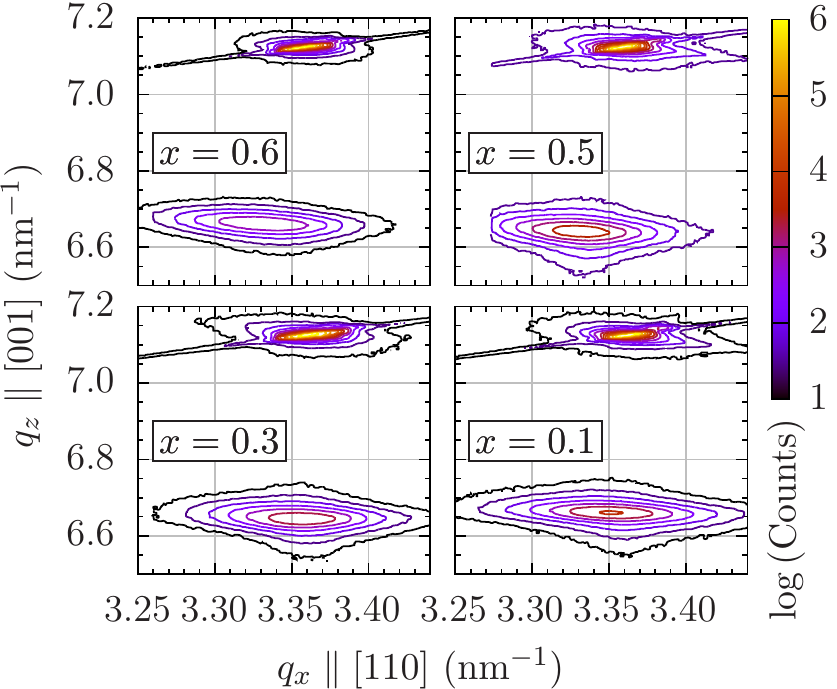}
  \caption{\label{fig:rsm} Reciprocal space maps around the MgO $[113]$ (MRA
  $[204]$) reflection from which we infer the in-plane lattice parameter of
  MRA\@. The samples are ordered in the sample surface plane.}
\end{figure}
The RSM data was collected around the (113) reflection of MgO and shown in
\figurename~\ref{fig:rsm}. The in-plane epitaxial relationship of MgO and MRA is
assumed to be MgO [100] parallel MRA [110], therefore we label the MRA peak as the
(204) reflection. We determined the in-plane ($a$) and out-of-plane lattice
parameters ($c$). The epitaxy improves with increasing Ru content and the films
are near-cubic, with a maximal tetragonal distortion $\left(
\frac{c-a}{a}\right)$ of \qty{1}{\percent} found for \mrx{0.39}. We summarize
the structural parameters of the samples in \tablename~\ref{tab:structure}.
\begin{table}
  \caption{\label{tab:structure}Crystal parameters of \mra{}. The lattice
    parameters in columns three and four are determined from reciprocal space
    mapping, see \figurename~\ref{fig:rsm}. The experimentally observed ratio
    of $S = F^2_{002}/F^2_{004}$ is compared to that calculated for a fully
    ordered structure with the stoichiometry deduced from the film density. Mn
    fully occupies the $4b$ position, and Ru most of the $8c$. The remaining $8c$
    is filled by Mn. Al occupies only $4a$.}
  \addtolength{\tabcolsep}{4pt}
  \sisetup{round-pad=false,round-mode=places,round-precision=2, table-alignment-mode = format}
\begin{tabular}{
	c 
	S[table-format = 1.2] 
	S[table-format = 3.2] 
	S[table-format = 3.2] 
	S[table-format = 1.2] 
	S[table-format = 1.2]}
\toprule
\# & $x$ & {$a\thinspace(\unit{\pm})$} & {$c\thinspace(\unit{\pm})$} & {$S$(exp)} & {$S$(th)} \\
\midrule
S08 &  0.0999330146719533 &  596.80 &  600.53 &  0.251374 &  0.262517 \\
S11 &  0.101857534914023 &  596.95 &  601.54 &  0.228655 &  0.262418 \\
S13 &  0.101857534914023 &  597.17 &  600.79 &  0.225299 &  0.262418 \\
S16 &  0.23144189788006 &  595.87 &  601.92 &  0.227388 &  0.252143 \\
S12 &  0.247479566563975 &  596.05 &  602.31 &  0.204882 &  0.250787 \\
S03 &  0.359743247351384 &  596.82 &  602.19 &  0.194581 &  0.241531 \\
S09 &  0.488044596822707 &  600.81 &  602.12 &  0.244588 &  0.230543 \\
S20 &  0.552195271558369 &  598 &  600 &  0.292826 &  0.225094 \\
S15 &  0.605440331588968 &  600.41 &  600.39 &  0.296349 &  0.220303 \\
S10 &  0.840873307868847 &  600.11 &  598.70 &  0.314845 &  0.198813 \\
\bottomrule
\end{tabular}

\end{table}
The degree of Ru order is estimated from the ratio $S =
F^2_{002}/F^2_{004}$. L2$_1$-ordered ($Fm\bar{3}m$) \mrx{0.0} and \mrx{1.0} have $S =
\numlist[round-mode = places, round-precision = 2]{0.333332;0.183622}$,
respectively. For the XA variants ($F\bar{4}3m$), both \mrx{0.0} (Ru on $4b$ and $4d$, Mn in
$4c$) and \mrx{1.0} (Mn on $4b$ and $4d$, Ru occupying $4c$) have 
$S \approx \num{0.02}$. 
Complete disorder among all elements in all positions suppresses the $(002)$
reflection completely and thus $S = 0$. We calculated the theoretical ratio $S$
using the stoichiometries inferred from the densities and an isotropic Debye-Waller
factor of \qty{0.3}{\angstrom\squared} for all atoms. We find
that for $x < 0.5$ the calculated and the observed $S$ agree reasonably
well, while for $x > 0.5$, the experimentally observed $S$ is
higher than predicted by the fully-ordered model indicating a degree of disorder. This
disordered phase is likely to retain the overall structure, but with increased
Debye-Waller factors. A much-improved agreement with the observed ratios for
$x = \numlist{0.55;0.61;0.84}$ is
obtained supposing the Debye-Waller factor is \qty{0.9}{\angstrom\squared}.

\subsection{\label{ssec:magnetics}Magnetism and magnetotransport properties}
\begin{figure}
  \includegraphics{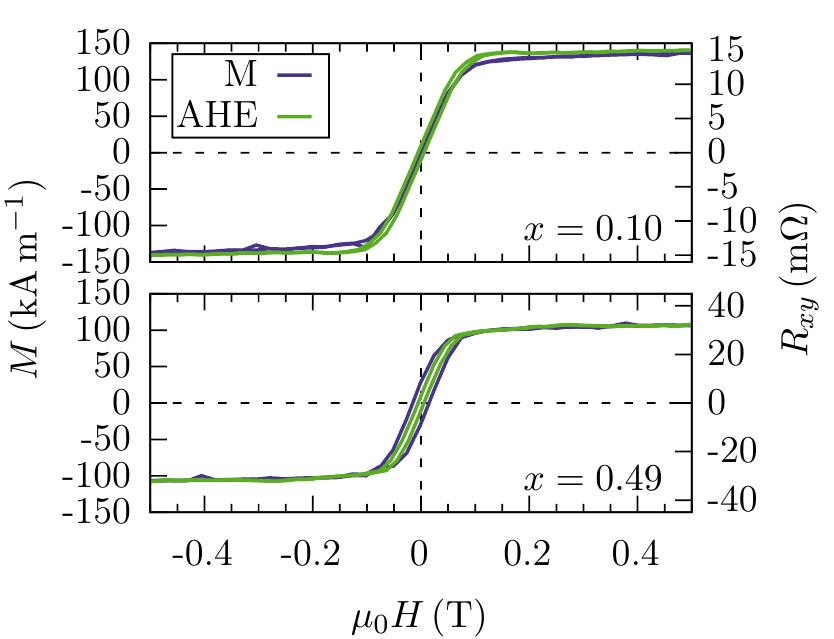}
  \caption{\label{fig:squid_ahe} Magnetometry and AHE recorded on two
    representative samples at \qty{296}{\kelvin}. The applied magnetic field is perpendicular to the sample
    surface. In the upper panel, the AHE loop is plotted as a function of
    $(- \mu_0 H)$ for easier comparison with the magnetometry data. Both
    techniques indicate easy plane shape anisotropy with the saturation field (i.e.\
    the anisotropy field) near-equal to $\mu_0 M_s$. The magnetometry data were
    corrected for the diamagnetism of the substrate by linear subtraction of
    the high field slope, as was the normal contribution to the transverse
    voltage.}
\end{figure}
We now turn to the magnetotransport properties of MRA\@. In
\figurename~\ref{fig:squid_ahe} we show magnetometry and anomalous Hall
effect (AHE) loops of two representative samples, where the diamagnetic
contribution from the substrate was subtracted as a linear slope in the
magnetometry data, as was the high-field contribution to the transverse
resistivity due to the normal Hall effect. 
The shape of the hysteresis loops recorded using the two techniques are
near-identical suggesting that the AHE reflects the net magnetisation of the
sample. Both magnetometry and AHE indicate weak easy-plane anisotropy, which is
a result of the sample shape. A small contribution from
perpendicular magnetocrystalline anisotropy due to the \qty{1}{\percent} tetragonal
distortion of the unit cell is present and therefore $\mu_0 H_{an} < \mu_0
M_s$. We measured hysteresis loops of a \mrx{0.1} sample at different
temperatures in order to extract $M_s$ and $H_{an}$.  From these we infer $K_1$
as a function of temperature. Assuming that $K_1(T) = K_1(0) \times \left (
M_s(T)/M_0 \right )^3$ we find $K_1(0) =
\qty{5.4}{\kilo\joule\per\metre\cubed}$.

Weiss mean field theory was used to fit the temperature-dependent magnetisation data
to obtain the exchange parameters and estimate the Curie temperature for $x =
0.14$. We take the Mn $4b$ moment to be \qty{2.66}{\mub} from the
DFT calculation (see \figurename~\ref{fig:mvsx}) giving a sublattice
magnetisation of \qty{451}{\kilo\ampere\per\metre}. The Mn $8c$ (now $4d$) is
then deduced from the difference between the net magnetisation at low
temperature and that of the Mn $4b$ and Ru $4c$ sublattices to be $\approx \qty{3.1}{\mub}$ per Mn or
\qty{92}{\kilo\ampere\per\metre}. The choice of the quantum number $S$ in the
Brillouin function of the localized mean field theory is always problematic in
a metal. It should somehow take account of the $d$-electron hybridisation with
Al and Ru in our alloy. A choice of $S = {}^5\!/\!_2$ for both Mn sites leads to
an excellent fit of the experimental data with three molecular field
coefficients $n_{\text{bb}}$, $n_{\text{bd}}$ and $n_{\text{dd}}$ and
coordination numbers $Z_{\text{bb}}$, $Z_{\text{bd}}$, $Z_{\text{dd}} = 12$,
$5.6$, and $4.8$ determined from the composition.  Our fit indicates that the
material orders just above room temperature at $T_C = $ \qty{353}{\kelvin}.
Heisenberg exchange energies were calculated to be $J_{\text{bb}}$,
$J_{\text{bd}}$, $J_{\text{dd}} = 0.96$, $-2.40$, and $1.33$
\unit{\milli\electronvolt} per
bond.  The ratio of the product of the coordination number and the Heisenberg
exchange energies is $J_{\text{bb}} Z_{\text{bb}} : J_{\text{bd}} Z_{\text{bd}}
: J_{\text{dd}} Z_{\text{dd}}$ equal to $10 : -10 : 2.4$. The exchange is a factor
of two smaller than for MRG, which explains the lower Curie temperature. The
net anisotropy constant is smaller than for MRG reflecting the shared crystal
structure and the similar degree of tetragonal distortion.

From the observed transverse ($\rho_{xy}$) and
longitudinal ($\rho_{xx}$) resistivity we determine the carrier concentration and
mobility assuming a single species of charge carriers. The results are
summarized in \tablename~\ref{tab:ahe}. The magnetic ordering temperature is
above room temperature for $x \lessapprox 0.6$.
\begin{table}
  \caption{\label{tab:ahe} Summary of \mra{} magnetic and transport properties.
    The magnetisation decreases monotonically with increasing $x$ for $x >
    0.1$.  We note that
    the sign of $\sigma_{xy}$ changes at $x \approx 0.25$. In the approximation
    of a single species of carriers contributing to the transport, the carrier
    concentration $n$ of the low-$x$ samples ($x \lessapprox 0.25$) is
    approximately twice that of the high-$x$ samples, while their mobility is
    half as great.}
  \sisetup{table-auto-round = true, per-mode = symbol, table-alignment-mode = format}
\begin{tabular}{
	c 
	S[table-format = 2.2] 
	S[table-format = 3.0] 
	S[table-format = 1.2, drop-exponent=true, exponent-mode=fixed, fixed-exponent=6] 
	S[table-format = +1.2, drop-exponent=true, exponent-mode=fixed, fixed-exponent=3] 
	S[table-format = 1.2, drop-exponent=true, exponent-mode=fixed, fixed-exponent=28] 
	S[table-format = 1.2, drop-exponent=true, exponent-mode=fixed, fixed-exponent=-3]} 
\toprule 
\# & {$x$} & {$M_s$} & {$\sigma_{xx}$} & {$\sigma_{xy}$} & {$n$} & {$\mu$} \\ 
& & {(\unit{\kilo\ampere\per\metre})} & {(\unit{\mega\siemens\per\metre})} & {(\unit{\kilo\siemens\per\metre})} & {($\unit{10^{28}\per\metre\cubed}$)} & {($\unit{\milli\volt\metre\per\second\squared}$)} \\ 
\midrule
S08 &  0.0999330146719533 &  137 &  400000 &  -713.6 &  3.686E+028 &  0.0000571 \\
S11 &  0.101857534914023 &  125 &  414937.7593361 &  -810.936450818684 &  2.1754E+028 &  0.0001003 \\
S13 &  0.101857534914023 &  150 &  400000 &  -620.8 &  2.0286E+028 &  0.0001037 \\
S12 &  0.247479566563975 &  135 &  423728.813559322 &  -1378.9141051422 &  2.7567E+028 &  0.0000808 \\
S16 &  0.23144189788006 &  135 &  346020.761245675 &  601.046443409442 &  1.2994E+028 &  0.00014 \\
S03 &  0.359743247351384 &  130 &  n/a &  n/a &  n/a &  n/a \\
S09 &  0.488044596822707 &  107 &  330033.00330033 &  926.924375605878 &  8.951E+027 &  0.0001939 \\
S20 &  0.552195271558369 &  100 &  n/a &  n/a &  n/a &  n/a \\
S15 &  0.605440331588968 &  76 &  350877.192982456 &  603.262542320714 &  9.915E+027 &  0.0001861 \\
S10 &  0.840873307868847 &  25 &  n/a &  n/a &  n/a &  n/a \\
\bottomrule
\end{tabular}

\end{table}

The sign of the transverse condictivity $\sigma_{xy} \approx
\rho_{xy}/\rho_{xx}^2$ changes at $x \approx 0.26$. Such a sign change is usually attributed to crossing
the compensation point where $M_s \approx 0$.
\begin{figure}
  \includegraphics{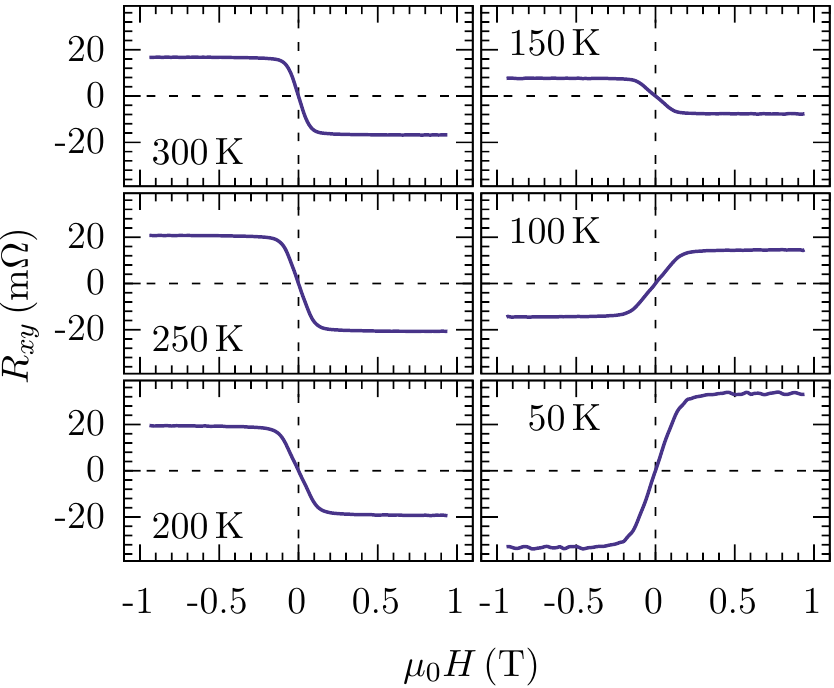}
  \caption{\label{fig:ahe-t} AHE recorded at
    \qtylist[list-units=single]{300;250;200;150;100;50}{\kelvin} for sample
    S08. The sign of $\sigma_{xy}$ changes between \qtylist{150;100}{\kelvin},
    but the anisotropy field remains near constant, indicating that this change
    is due to a change in the nature (concentration and mobility) of the
    predominant charge carrier rather than crossing a temperature where magnetic
    compensation occurs.}
\end{figure}
In \figurename~\ref{fig:ahe-t} we plot the transverse resistance measured for
\mrx{0.13} at \qtylist[list-units=single]{300;250;200;150;100;50}{\kelvin}. As
when changing $x$, its sign changes between \qtyrange{150}{100}{\kelvin}. The
anisotropy field, however, changes monotonically from \qtyrange{0.1}{0.2}{\tesla}
between room temperature and \qty{50}{\kelvin} without any divergence
indicating an $M_s$ close to zero. We therefore attribute the sign
change to a temperature-dependent filling of two or more pockets in the band
structure near the Fermi energy, in agreement with the change of carrier
concentration and their mobility shown in \tablename~\ref{tab:ahe}. The change
in the anisotropy field due to shape anisotropy suggests the saturation
magnetisation increases by $\approx \qty{0.9}{\mub\per\formulaunit}$ between
\qtyrange{300}{50}{\kelvin}, in agreement with low-temperature magnetometry
measured on other samples (not shown).

\begin{figure}
  \includegraphics{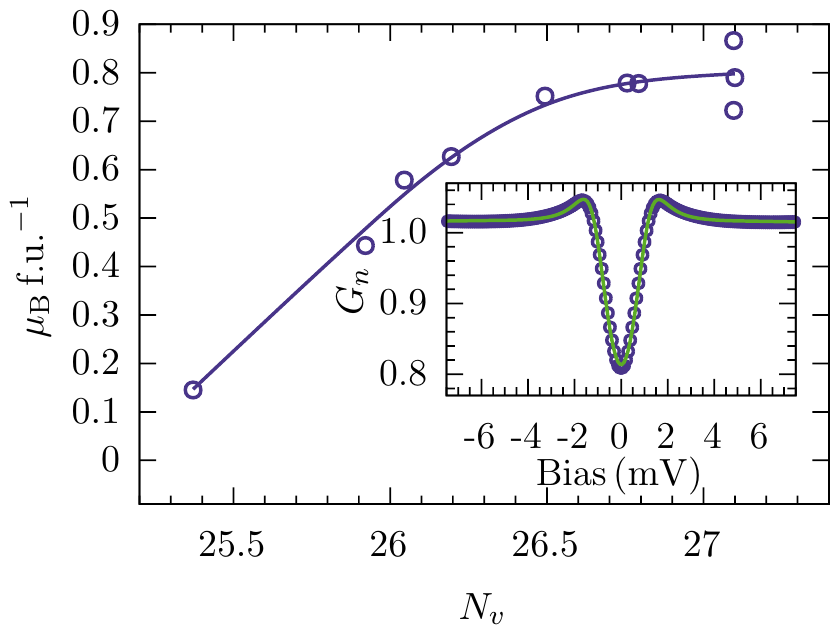}
  \caption{\label{fig:m_and_pcar} Magnetisation of \mra{} as a function of number of
    valence electrons ($N_v$). The solid line is a guide to the eye. For
    $N_v \lessapprox 26.5$ the magnetisation increases by
    $\approx \qty{0.6}{\mub\per\formulaunit\per\elementarycharge}$. The insert
    shows a point contact Andreev reflection (PCAR) spectrum for $x = 0.25$. The spin
  polarisation inferred from the fit (solid line) is $P = \qty{52}{\percent}$.}
\end{figure}
The results indicate that \mra{} is a ferrimagnet where Ru in
$8c$ and Mn in $4b$ form antiparallel ferromagnetic sublattices. No
compensation composition is found in
the range of $x$ we have investigated, nor is there a compensation temperature
due to the different local environments of Mn and Ru. Two (or more) species of charge
carriers with opposite spin dominate the Fermi level, and their
concentrations and mobilities differ by around a factor of two. \mra{} is 
not a half metal. In \figurename~\ref{fig:m_and_pcar} we show the saturation
magnetisation at \qty{300}{\kelvin} as a function of the number of valence
electrons $N_v$. The dependence is non-linear, however for low $N_v < 26.3$, the
magnetisation increases with
\qty{0.6}{\mub\per\formulaunit\per\elementarycharge}, suggesting a spin
polarisation $P$ around \qty{60}{\percent} in this region. The insert in
\figurename~\ref{fig:m_and_pcar} shows point-contact Andreev reflection
recorded on \mrx{0.4}. We find the spin polarisation $P = \qty{52}{\percent}$
in good agreement with that inferred from the plot of $M$ against $N_v$. 

\subsection{\label{ssec:dft}Comparison with density functional theory}
Finally we compare our results with \emph{ab initio} calculated magnetisation and density of
states. We used a 16-atom cell for all our calculations, and relaxed both
lattice parameters ($a \approx \qty{600}{\pico\metre}$) and magnetic moments, but
kept the individual atoms at their assigned Wychoff positions in agreement with
the structural model inferred from the analysis of the x-ray data above.
\begin{figure}
  \includegraphics{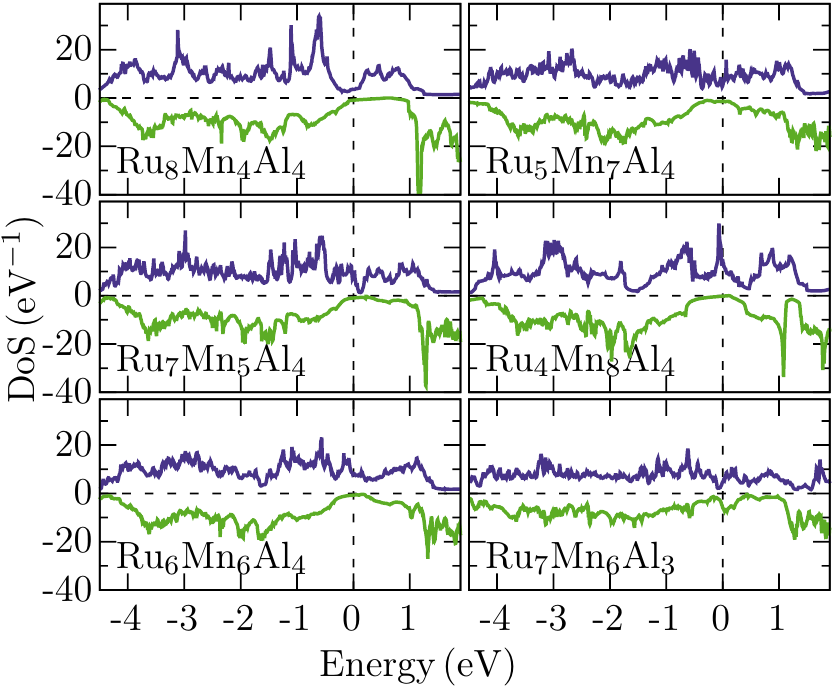}
  \caption{\label{fig:dos} Density of states for \mra{} with $x$ ranging from
  \numrange{0}{1}. The composition $\text{Ru}_7\text{Mn}_6\text{Al}_3$ was selected
as it corresponds closely to the experimental sample S16. It illustrates the
change in the density of states due to the under stoichiometry in Al, although
neither the spin polarisation ($P \approx \qty{50}{\percent}$) nor the magnetic
moment ($M \approx \qty{1.5}{\mub\per\formulaunit}$) changes significantly
compred to the ordered version ($\text{Ru}_7\text{Mn}_5\text{Al}_4$). }
\end{figure}
In \figurename~\ref{fig:dos} we show the density of states (DoS) around the Fermi
level for a series of \mra{} samples with varying $x$. The label on each panel
corresponds to the content of the 16-atom cell used for calculations, with Ru
occupying the $8c$ and Mn and Al the $4b$ and $4a$ positions, respectively.
For \mrx{0}, in particular, there is a gap in the spin down DoS, but its onset is
almost \qty{1}{\electronvolt} above the Fermi level. With decreasing Ru (and
increasing Mn) the gap moves closer towards the Fermi level, but is
simultaneously destroyed by the creation of supplementary states in the gap.
None of the compositions are half metallic, although perfectly ordered \mrx{1}
in the top left panel of \figurename~\ref{fig:dos} comes close. We have seen
above that this composition disorders during growth.  The bottom right panel of
\figurename~\ref{fig:dos} illustrates the effect of Al substoichiometry.
From these figures we estimate the spin polarisation of
MRA to be around \qty{50}{\percent}, in good agreement with the experimental
results.

\begin{figure}
  \includegraphics{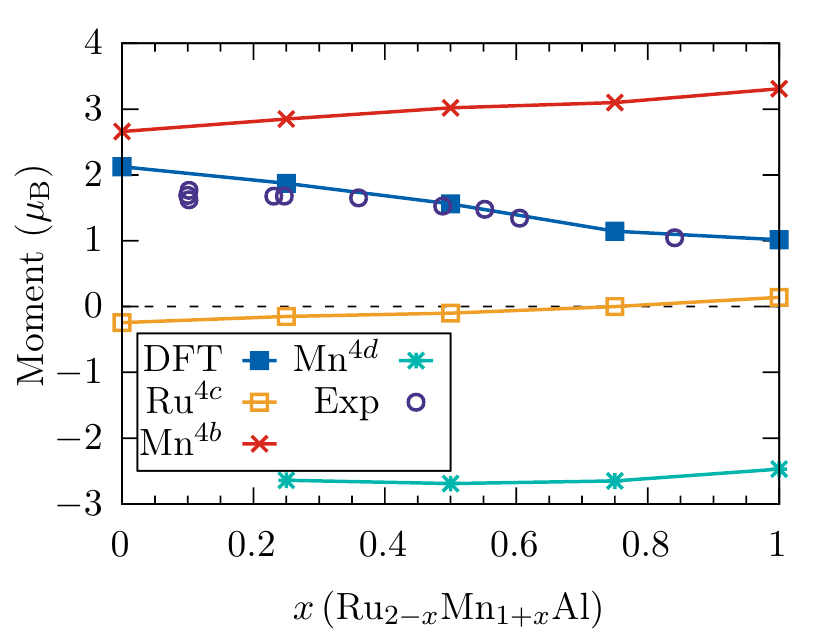}
  \caption{\label{fig:mvsx} Experimental and calculated net and sublattice
  moments for $x$ in the range \numrange{0}{1}. The net calculated and
  experimental
  moments are given per formula unit. Note that the experimental values
  have been shifted by \qty{0.9}{\mub\per\formulaunit} to account for the higher
  magnetisation at $T = \qty{0}{\kelvin}$.}
\end{figure}

We also show, in \figurename~\ref{fig:mvsx}, the site- and ion-resolved
magnetic moments obtained by theory. The net moment decreases from
$\approx \qtyrange{2}{1}{\mub\per\formulaunit}$ for $x$ in the range
\numrange{0}{1}.
Mn in the $4b$ position carries a strong moment $\approx \qty{3}{\mub}$ that
increases with increasing $x$, suggesting increased localisation of the $3d$
bands. As $x$ is raised above zero, some Mn fills the $8c$ (now $4d$) position
and it is coupled antiferromagnetically to Mn in $4b$, but ferromagnetically to
the remaining Ru (in $8c$/$4c$). Ru carries a small moment in all compositions,
but interestingly changes sign at $x \approx 0.75$. In the figure we also plot
the experimentally observed magnetisation, offset by
\qty{0.9}{\mub\per\formulaunit} to account for the decrease of the
magnetisation between $T = \qtylist{0;300}{\kelvin}$ (see above).

\section{\label{sec:conclusions}Conclusions} 
We find that \mra{} crystallises on MgO and MgAl$_2$O$_4$ substrates in a
tetragonally-distorted
near-cubic $Fm\bar{3}m$ crystal structure. For low values of $x$ ($< 0.5$), the
structure remains essentially ordered with Ru in the $8c$ position and Mn and
Al in $4b$ and $4a$, respectively, with the additional Mn filling the sites
vacated by Ru. As $x$ is increased further, the crystal structure
evolves towards $F\bar{4}3m$, although order between the $4c$ and $4d$ sites
does not occur, and additionally at low $x$, the $4a$-$4b$ order is perturbed. All
compositions are magnetically ordered and exhibit a high spin
polarisation of around \qty{50}{\percent}. \emph{Ab initio} calculations
agree well with the observed crystal structure and magnetic mode. MRA is
ferrimagnetic, with Mn in the $4b$ position coupling antiferromagnetically to
Ru in $8c$. Two or more species of charge carriers close to or at
the Fermi level account for the reduction of the spin polarisation and
the change in the sign of the Hall conductivity with $x$ and T. In contrast to MRG, where the spin
polarisation as inferred from the Slater-Pauling
plot\cite{Siewierska2021_PRB} is close to $P = \qty{100}{\percent}$, the
anomalous Hall angle ($\sigma_{xy}/\sigma_{xx}$) of MRA is only $\sim
\qty{2}{\permille}$, more than an order of magnitude less than for MRG despite
a higher concentration of Ru.

When $x$ is increased above
$\approx 0.6$, MRA no longer exhibits AHE at room temperature although the crystal structure is
increasingly similar to that of MRG\@. We therefore conclude that Ga,
and not Ru, is promoting high conduction band spin orbit coupling in MRG and
helps to move the Fermi level into the spin gap by allowing Ga-Al antisites to
form.\cite{Zic.2016} It can be argued that MRA retains inversion symmetry due
to disorder, even when $x$ is above $\sim 0.6$. MRG has no centre of inversion
because Ru and Mn order on the $4d$ and $4c$ positions, and a
topological contribution to the transverse conductivity is
allowed by symmetry.

At $x = 0$, the Mn-Mn distance is \qty{600}{\pico\metre}, and the Mn orders
ferromagnetically. The moment carried by Ru is small, and it is
unlikely that the magnetic order is due to the Ru-Mn AFM interaction. The
question of the nature of this long-distance exchange interaction arises.
Several other Heusler alloys (Rh$_2$Mn(Pb;Sn;Ge) and
Cu$_2$Mn(In;Sn;Al))\cite{hpjWijn1991,PhysRevB.14.4131} where only one out of four atoms per formula is
manganese order ferromagnetically above \qty{300}{\kelvin}, and
Au$_4$Mn\cite{he2022} with
a shorter first neighbour Mn-Mn distance of \qty{404}{\pico\metre} has a Curie point at
\qty{390}{\kelvin}. There, all the exchange interactions up to
the tenth nearest-neighbours at \qty{1000}{\pico\metre} are either ferromagnetic or zero. Since different
elements are present in these dilute Mn-based ferromagnets, it is likely that
long-range ferromagnetic exchange is mediated by a $p$-band. The Mn-Mn exchange
(in Au$_4$Mn) only becomes oscillatory and RKKY-like at Mn-Mn distances exceeding
\qty{1040}{\pico\metre}.

In summary, \mra{} is a ferrimagnetic Heusler alloy with high spin
polarisation and magnetic ordering at room temperature. The absence of
signifiant spin-orbit coupling precludes perpendicular magnetic anisotropy as
well as any notable magneto-resistive effects. This contrast with MRG where Al is
replaced by Ga, is likely due to a combination of hybridisation of the Mn $d$ and $s$
states with Ga $p$ bands, and the presence, or absence, of a centre of inversion.

\begin{acknowledgments}
  K.E.S.\@ and J.M.D.C.\@ acknowledge funding through Irish Research Council
  under Grant GOIPG /2016/308 and in part by the Science Foundation Ireland
  under Grants 12/RC/2278 and 16/IA/4534. B.S.\@ acknowledges funding from
  the Irish Research Council Grant EPSPG/2020/106G.A. A.J., G.A., Z.G., and
  K.R.\@ acknowledge funding from `TRANSPIRE' FET Open Programme, H2020.
  N.T.\@ was supported by the European Union's Horizon 2020 Research and
  Innovation Programme under Marie Sk\l{}odowska-Curie EDGE Grant 713567.
  H.K.\@ and M.V.\@  were funded through the ZEMS contract, Science Foundation
  Ireland, grant 16/IA/4534.
\end{acknowledgments}

\bibliographystyle{apsrev4-2}
\bibliography{bib}

\end{document}